\documentclass[%
aps,
prapplied,
twocolumn,
groupedaddress,
superscriptaddress,
amsmath,amssymb,
]{revtex4-2}

\usepackage{graphicx}
\usepackage{dcolumn}
\usepackage{bm}
\usepackage{hyperref}
\usepackage{color}

\begin{document}

\title{Diamond quantum magnetometer with dc sensitivity of < 10 pT Hz$^{-1/2}$ toward measurement of biomagnetic field}


\author{N. Sekiguchi}
\email[]{sekiguchi.n.ac@m.titech.ac.jp}
\affiliation{Department of Electrical and Electronic Engineering, Tokyo Institute of Technology, Meguro, Tokyo 152-8550, Japan}

\author{M. Fushimi}
\affiliation{Department of Bioengineering, The University of Tokyo, Bunkyo, Tokyo 113-8656, Japan}

\author{A. Yoshimura}
\affiliation{Department of Electrical and Electronic Engineering, Tokyo Institute of Technology, Meguro, Tokyo 152-8550, Japan}

\author{C. Shinei}
\affiliation{Research Center for Electronic and Optical Materials, National Institute for Materials Science, Tsukuba, Ibaraki 305-0044, Japan}

\author{M. Miyakawa}
\author{T. Taniguchi}
\affiliation{Research Center for Materials Nanoarchitectonics, National Institute for Materials Science, Tsukuba, Ibaraki 305-0044, Japan}
\author{T. Teraji}
\affiliation{Research Center for Electronic and Optical Materials, National Institute for Materials Science, Tsukuba, Ibaraki 305-0044, Japan}

\author{H. Abe}
\author{S. Onoda}
\author{T. Ohshima}
\affiliation{Takasaki Institute for Advanced Quantum Science, National Institutes for Quantum Science and Technology, Takasaki, Gunma 370-1292, Japan}

\author{M. Hatano}
\affiliation{Department of Electrical and Electronic Engineering, Tokyo Institute of Technology, Meguro, Tokyo 152-8550, Japan}

\author{M. Sekino}
\affiliation{Department of Bioengineering, The University of Tokyo, Bunkyo, Tokyo 113-8656, Japan}

\author{T. Iwasaki}
\affiliation{Department of Electrical and Electronic Engineering, Tokyo Institute of Technology, Meguro, Tokyo 152-8550, Japan}

\date{\today}

\begin{abstract}
We present a sensitive diamond quantum sensor with a magnetic field sensitivity of $9.4 \pm 0.1~\mathrm{pT/\sqrt{Hz}}$ in a near-dc frequency range of 5 to 100~Hz.
This sensor is based on the continuous-wave optically detected magnetic resonance of an ensemble of nitrogen--vacancy centers along the [111] direction in a diamond (111) single crystal.
The long $T_{2}^{\ast} \sim 2~\mathrm{\mu s}$ in our diamond and the reduced intensity noise in laser-induced fluorescence result in remarkable sensitivity among diamond quantum sensors.
Based on an Allan deviation analysis, we demonstrate that a sub-picotesla field of 0.3~pT is detectable by interrogating the magnetic field for a few thousand seconds.
The sensor head is compatible with various practical applications and allows a minimum measurement distance of about 1~mm from the sensing region.
The proposed sensor facilitates the practical application of diamond quantum sensors.
\end{abstract}

\maketitle

\section{%
\label{sec: intro}
Introduction
}

The biomedical applications of quantum sensors have been studied for over a decade \cite{Asl23}.
The realization of magnetoencephalography (MEG) under ambient conditions is a major goal (conventional MEG requires a magnetically shielded room).
In addition to clinical diagnosis \cite{Har18, Uhl18}, ambient-condition MEG can be used for daily diagnosis, brain-machine interfaces \cite{Fuk18, Rat19}, and fundamental research on brain function \cite{Bot18, Hil19, Gro19, Bai17}.
A quantum magnetometer that uses nitrogen--vacancy (NV) centers in diamond is a candidate for realizing ambient-condition MEG given that it can be operated with high sensitivity at room temperature in an ambient magnetic field \cite{Tay08, Aco09, Bar16, Zha21, Shi22, Gra23, Fes20, Bal09, Wol15}.
A spatial resolution on the millimeter scale or below, far better than the centimeter-scale resolution of conventional MEG \cite{Uhl18}, is expected for a diamond quantum magnetometer \cite{Ara22}.

Magnetometry based on continuous-wave optically detected magnetic resonance (CW-ODMR) is the most widely used method for measuring magnetic fields using NV centers \cite{Tay08, Aco09, Zha21, Gra23, Shi22, Bar16, Fes20}.
In this method, a microwave (MW) field continuously drives the magnetic resonance of the NV center spin and the spin state is continuously read out as the intensity of the laser-induced fluorescence from the NV center.
Compared with other methods based on pulsed MWs and/or light \cite{Bal09, Zha21, Wol15, Bar23}, the CW-ODMR method has a simpler experimental setup and is easier to apply to actual measurements. 
Millimeter-scale magnetocardiography \cite{Ara22} has been realized using the CW-ODMR method. 
However, to realize MEG, measurement of an encephalomagnetic field requires exceptional sensitivity (on the order of $\mathrm{pT/\sqrt{Hz}}$).
The frequency of a clinically relevant encephalomagnetic field ranges from nearly dc to $\sim 100~{\mathrm Hz}$ \cite{Uhl18, Har18, Bai17}.
Reported sensitivities are worse than required in this frequency range.
For example, sensitivities of around $20$ to $30~\mathrm{pT/\sqrt{Hz}}$ \cite{Zha21, Gra23} have been demonstrated.
In addition, a sensitivity of $15~\mathrm{pT\sqrt{Hz}}$ in a higher frequency range (80~Hz to 3.6~kHz) has been reported \cite{Bar16}.
A short standoff distance from field generating sources in the brain is also required given that the decay of an encephalomagnetic field is inversely proportional to the square of the distance \cite{Har18}.
Therefore, for biomedical applications, the sensitivity in the near-dc frequency range of a diamond quantum magnetometer that can closely approach the target object must be improved.

Here, we develop a CW-ODMR-based diamond magnetometer for practical applications (e.g., MEG of a living animal).
The sensor head of the magnetometer was designed to approach the target object to a distance of about 1~mm with a sensing volume of $0.03~\mathrm{mm^{3}}$.
By carefully tuning the experimental conditions and using a high-quality diamond, we achieved a record-breaking sensitivity of $9.4 \pm 0.1~\mathrm{pT/\sqrt{Hz}}$ in a near-dc frequency range of 5 to 100~Hz.
Based on the Allan deviation, the minimum detectable field was found to be 8.5 and 0.3~pT for measurement periods of 1 second and several thousand seconds, respectively.

\section{%
\label{sec: setup}
Experimental setup
}

\subsection{%
\label{ssec: head}
Sensor head
}

In this work, we synthesized a single-crystalline diamond using a high-pressure--high-temperature (HPHT) method with a $^{12}$C isotopically enriched carbon source.
The reduced concentration of $^{13}$C was about 500 ppm.
The amount of titanium in the metal solvent in the HPHT synthesis was adjusted to control the initial concentration of neutral substituted nitrogen ($\mathrm{N_{s}^{0}}$) in the diamond crystal \cite{Miy22}.
The initial $[\mathrm{N_{s}^{0}}]$ was estimated to be 5.6~ppm using electron spin resonance.
The origin of nitrogen in diamond crystals seems to be impurities introduced from the source material, solvent, or pressure transmitting medium during the growth process. 
Since this nitrogen is of natural origin, the isotope abundance of $\mathrm{N_s^0}$ is the same as the natural abundance ($^{14}$N, 99.6\%; $^{15}$N, 0.4\%).
After this HPHT synthesis, a piece of the crystal was cut out parallel to the (111) crystal plane.
The dimensions of this diamond sample were approximately $1~\mathrm{mm} \times 0.7~\mathrm{mm}$ in area and 0.4~mm in thickness.
Negatively charged NV ($\mathrm{NV^{-}}$) centers were then produced using electron beam irradiation followed by annealing at $1000~\mathrm{C^{\circ}}$ for 2 hours in vacuum.
The energy and total fluence of the irradiation were 2.0~MeV and $5\times 10^{17}~\mathrm{cm^{-2}}$, respectively.
The concentrations of the produced $\mathrm{NV^{-}}$ and residual $\mathrm{N_s^{0}}$ were estimated to be 1.2 and 2.3 ppm, respectively, using electron spin resonance \cite{Shi22a}.
A full width at half maximum of 0.19~MHz for the CW-ODMR peak was experimentally measured independent of this work. This linewidth indicates a long dephasing time of $T_2^{\ast} \sim 2~\mathrm{\mu s}$.

\begin{figure}
\includegraphics{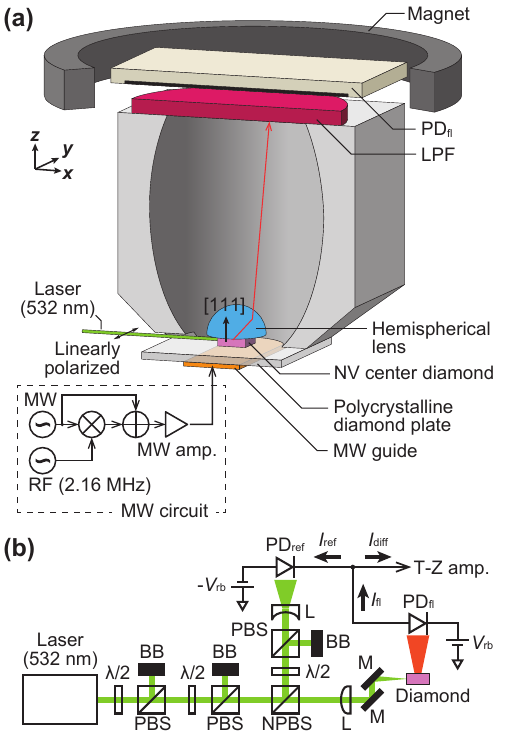}
\caption{\label{fig: setup}
 	Experimental setup (not to scale).
	(a) Sensor head design and MW circuit diagram.
	$\mathrm{PD_{fl}}$: fluorescence photodiode; 
	LPF: long-pass filter.
	(b) Optical setup. 
	$\lambda / 2$: halfwave plate; 
	PBS: polarizing beam splitter; 
	NPBS: non-polarizing beam splitter; 
	L: lens; 
	M: mirror; 
	$\mathrm{PD_{ref}}$: reference photodiode; 
	BB: beam block.
}
\end{figure}

The conceptual design of our sensor head is shown in Fig.~\ref{fig: setup}(a).
This sensor head was designed to closely approach the head of a living animal and measure the encephalomagnetic field along the $z$ axis by an ensemble of NV centers oriented to the surface-normal [111] direction parallel to the $z$ axis.
The sensor head components described in this section (see below) were integrated using plastic and aluminum holders.
Hence, the sensor head can be freely moved as a unit and easily positioned close to the target object.

The diamond containing NV centers was attached by a high-thermal-conductivity glue to a polycrystalline diamond plate ($10\times 10 \times 0.5~\mathrm{mm^{3}}$)  in order to dissipate the heat due to laser illumination.
The other side of the polycrystalline diamond plate had a current flow guide for MWs.
The MW guide was made of thin copper film; the distance between the lower side of the MW guide and the excited NV centers was 0.8~mm.
A bias magnetic field of 0.9~mT along the $z$ axis was applied by a ring samarium-cobalt magnet.

We used a hemispherical lens with a high refractive index of 2.0 to enhance the collection efficiency of the laser-induced fluorescence from the NV center ensemble \cite{Shi22}.
The fluorescence collection efficiency from the diamond surface facing the lens [top surface in Fig.~\ref{fig: setup}(a)] was assumed to be as high as about 56\% based on a previously reported numerical calculation~\cite{Shi22} for a similar setup.
The fluorescence that was not emitted from this surface was considered to be emitted mainly from the side faces due to the high refractive index (2.4) of the diamond \cite{Le12}.
Some of the fluorescence from the side faces of the diamond was collected by the lens since the lens diameter (4~mm) was larger than the size of the diamond.
Fluorescence was also collected by an elliptically shaped reflective inner surface of an aluminum block.
Stray green light and part of the fluorescence from neutrally charged NV ($\mathrm{NV^{0}}$) centers were filtered out by a long-pass filter with a cut-on wavelength of 633~nm. The transmitted fluorescence was detected by a reverse-biased photodiode ($\mathrm{PD_{fl}}$).

\subsection{%
\label{ssec: meas setup}
CW-ODMR measurement setup
}

The NV ensemble was excited by a green laser at 532~nm from a side face of the diamond, as shown in Fig.~\ref{fig: setup}(a).
Figure~\ref{fig: setup}(b) shows the optical setup.
A laser beam with a diameter of about 3~mm was focused by a lens with a focal length of 300~mm.
The beam diameter at the diamond was estimated to be $200~\mathrm{\mu m}$.
The excitation volume was estimated to be $0.03~\mathrm{mm^{3}}$.
The laser light was linearly polarized along the $y$ axis, which is perpendicular to the chosen NV orientation.
The fluorescence photocurrent $I_{\mathrm{fl}}=6.6~\mathrm{mA}$ was observed at an incident light power of 0.39~W, which corresponds to a detected fluorescence power of about 13~mW.

The noise in the fluorescence due to the intensity fluctuation of the incident laser was reduced using a balanced detection technique.
The reference light, which was picked up by a non-polarizing beam splitter, was detected by a reverse-biased photodiode ($\mathrm{PD_{ref}}$).
In this work, we connected the anode of $\mathrm{PD_{fl}}$ to the cathode of $\mathrm{PD_{ref}}$ to obtain the difference between their photocurrents, $I_{\mathrm{fl}}$ and $I_{\mathrm{ref}}$, respectively.
The difference photocurrent $I_{\mathrm{diff}}$ was amplified by a lab-built transimpedance amplifier with a gain of 10~kV/A.
The power of the reference light was finely adjusted using a halfwave plate and a polarizing beam splitter to achieve a high reduction rate for the intensity noise.
The polarization fluctuation of the laser was converted into an intensity fluctuation by a polarization beam splitter just after the laser.
The beam diameter at $\mathrm{PD_{ref}}$ was expanded by a lens to balance the nonlinear response of the photodiode with that of $\mathrm{PD_{fl}}$, since the nonlinear response depends on spot size \cite{Sch04}.

The magnetic resonance between the ground states $|0\rangle$ and $|-1\rangle$ was driven by applying an MW current to the MW guide.
To enhance the amplitude of a CW-ODMR peak, we simultaneously drove the three transitions associated with the hyperfine spin state by three-tone MWs \cite{Bar16}, which was generated by mixing radio-frequency (RF) waves at 2.16~MHz with MWs and summing the mixed waves with bypassed MWs.
In this work, the enhancement factor for the peak amplitude was about 2.5.

We adopted lock-in detection, achieved by sinusoidally modulating the MW frequency, to avoid large residual noise at low frequencies.
The amplified difference photocurrent was fed into a lock-in amplifier and demodulated with the modulation frequency as the reference.
The 3-dB cutoff frequency of the low-pass filter in the lock-in amplifier was 149.4~Hz, which corresponds to a noise-equivalent-power bandwidth of 168.8~Hz.
The demodulated output was recorded on a computer via an analog-to-digital converter.

The sensor head and optical setup were inside a room that was shielded from magnetic fields by three permalloy layers to reduce environmental field fluctuations.
The total shielding factor of this room was about $2\times 10^{-4}$ at 1~Hz and about $1\times 10^{-5}$ at 10~Hz.
An additional permalloy box was placed around the sensor head.
The front face of the shield box remained open to introduce the incident laser and the target object.

\section{%
\label{sec: results}
Results
}

\subsection{
\label{ssec: odmr}
CW-ODMR measurement}

\begin{figure}
\includegraphics{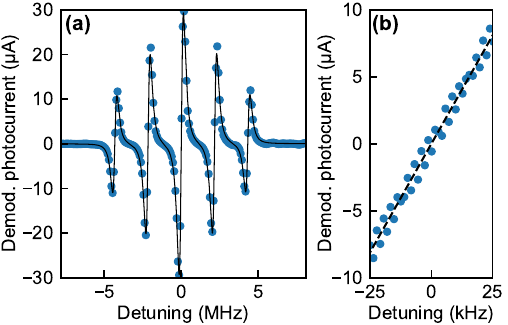}
\caption{\label{fig: odmr}
 	Lock-in CW-ODMR spectrum (a) over hyperfine manifold and (b) in near-resonant region of central peak.
	Measured demodulated photocurrent is shown by filled circles.
	The solid line in (a) represents the fitted curve obtained from the summation of five derivative Lorentzian functions.
	The linear function shown by the dashed line in (b) was fitted to the near-resonant data to obtain a zero-crossing slope.
}
\end{figure}

Figure~\ref{fig: odmr}(a) shows a CW-ODMR spectrum of the ensemble of [111]-oriented NV centers.
The vertical axis in the figure is the demodulated signal $\tilde{I}$ in the photocurrent, which was calculated using the gains at the transimpedance and lock-in amplifiers.
The horizontal axis is the detuning $\delta$ from the resonance frequency of the central peak at which the three hyperfine spin states were simultaneously driven.
The fluorescence photocurrent $I_{\mathrm{fl}}$ at a far-detuned MW frequency was $I_{\mathrm{fl}} = 6.6~\mathrm{mA}$.
In this measurement, the frequency and depth of the modulation were 6.2~kHz and 160~kHz, respectively.
We found that a modulation frequency of 3 to 7~kHz yielded a low-noise output.
The modulation frequency was finely tuned within this range on each day of the experiment because the frequencies of some noise peaks in the intensity noise slightly shifted over time.
The modulation frequency of higher than several kilohertz caused a decrease in the amplitude of the CW-ODMR peaks, probably because the dynamics of the population in the ground states was slower than the modulation at several kilohertz for the relatively low power of $< 1$~W and large spot size of 200~$\mathrm{\mu m}$ in diameter.
The MW and RF wave power was tuned to yield a maximum zero-crossing slope at the central peak.
The black solid curve is fitted to the measured data using the summation of five derivative Lorentzian functions.
The corresponding full width at half maximum of the derivative Lorentzian function was about 0.48~MHz .
This linewidth is greater than the inhomogeneous broadening of the CW-ODMR peak due to the inhomogeneity in the bias magnetic field, which was estimated to be approximately 0.2~MHz.

To determine the zero-crossing slope, we measured a CW-ODMR spectrum at the near-resonance region of the central peak, as shown in Fig.~\ref{fig: odmr}(b).
The demodulated photocurrent linearly depends on the detuning in this region.
The zero-crossing slope $d\tilde{I}/d\delta$ was measured to be $324~\mathrm{pA/Hz}$ by fitting the data with a linear function, as shown by the black dashed line.
This slope corresponds to the photocurrent response to magnetic field variation as $(d\tilde{I}/d\delta) / \gamma_{\mathrm{e}} = 9.06~\mathrm{A/T}$, where $\gamma_{\mathrm{e}} = 28.0~\mathrm{GHz/T}$ is the gyromagnetic ratio for an NV center.

\subsection{%
\label{ssec: balance}
Reduction in intensity noise
}

The reduction rate for the intensity noise was estimated at $I_{\mathrm{fl}}=25~\mathrm{mA}$.
We measured the standard deviations of $\tilde{I}$ with and without the reference light to be $3.0~\mathrm{nA}$ and $130~\mathrm{nA}$, respectively.
Here, the MW source was switched off to isolate the sensor from the noise associated with the environmental magnetic field.
The relative intensity noise in the incident light was roughly estimated to be $10\mathrm{log_{10}}\left( \frac{130~\mathrm{nA}^{2}}{168.8~\mathrm{Hz} \times 25~\mathrm{mA}^{2}} \right)= -130~\mathrm{dBc/Hz}$ at a modulation frequency of 6.2~kHz.
The photon shot noise with and without the reference light was calculated to be $1.6~\mathrm{nA}$ and $1.2~\mathrm{nA}$, respectively.
The details of the photon shot noise calculation are described in Sec.~\ref{ssec: phc depend}.
We obtained the following reduction rate for the fluorescence intensity noise: 
\begin{equation*}
\sqrt{
\frac{
3.0~\mathrm{nA}^{2} - 1.6~\mathrm{nA}^{2}
}{
130~\mathrm{nA}^{2} - 1.2~\mathrm{nA}^{2} 
}
}
= 1.9\times 10^{-2}.
\end{equation*}
We found that this ``red--green'' balance detection exhibited a similar reduction rate to that for the ``green--green'' balanced detection with the reference and incident lights.

\subsection{%
\label{ssec: phc depend}
Photocurrent dependence of noise
}

\begin{figure}
\includegraphics{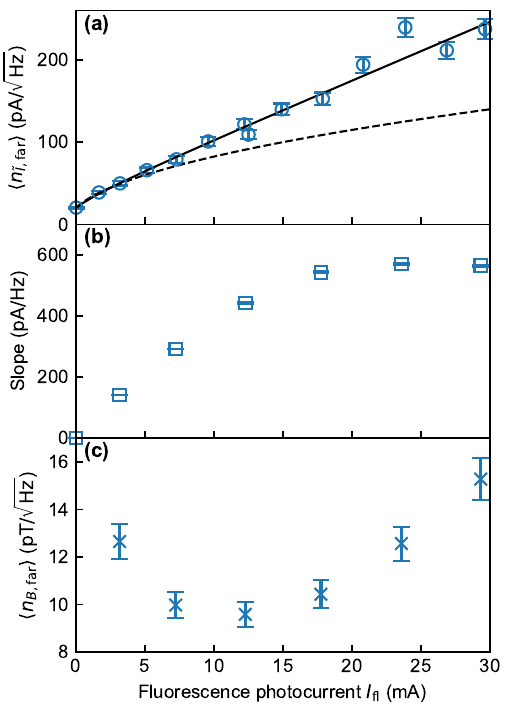}
\caption{\label{fig: noise}
 	Fluorescence photocurrent dependence of
	(a) floor of noise spectral density of $\tilde{I}$, 
	(b) zero-crossing slope $d\tilde{I}/d\delta$, and 
	(c) estimated floor of equivalent magnetic field noise spectral density.
}
\end{figure}

We analyzed the noise components (photon shot noise, fluorescence intensity noise, and electrical noise) of the detectors and circuits by measuring their dependence on the fluorescence photocurrent $I_{\mathrm{fl}}$.
The demodulated photocurrent $\tilde{I}$ was recorded for 5~s and Fourier-transformed to provide a single-sided noise amplitude spectral density $n_{\tilde{I}}$. 
To evaluate the noise $n_{\tilde{I}, \mathrm{far}}$ without influence from environmental magnetic field noise, the analysis was performed with an MW carrier frequency of 2.4~GHz, which was far-detuned from the resonance.
We observed no excess noise due to the application of the far-detuned MWs.
The noise density $n_{\tilde{I}, \mathrm{far}}$ was almost flat up to the cutoff frequency of the lock-in amplifier.
The average $\langle n_{\tilde{I}, \mathrm{far}} \rangle$ of the noise density within the 100-Hz bandwidth was taken as a measure of the intrinsic noise of our diamond sensor at a given $I_{\mathrm{fl}}$.
The dependence on $I_{\mathrm{fl}}$ of $\langle n_{\tilde{I}, \mathrm{far}} \rangle$ is shown in Fig.~\ref{fig: noise}(a).
Here, we varied the incident laser power using a halfwave plate and a polarizing beam splitter just before the non-polarizing beam splitter.
In the figure, the measured $\langle n_{\tilde{I}, \mathrm{far}} \rangle$ is represented by open circles.
The relative uncertainty in the data, shown as error bars, was independently evaluated to be 5\%.
$\langle n_{\tilde{I}, \mathrm{far}} \rangle$ at $I_{\mathrm{fl}}=0$ represents the electrical noise density  $\langle n_{\tilde{I}, \mathrm{elec}} \rangle$ and was measured to be $20~\mathrm{pA/\sqrt{Hz}}$ by blocking the laser beam before the non-polarizing beam splitter.

We fitted the noise model in Eq.~(\ref{eq: noise model}) to the data.
\begin{equation}
\label{eq: noise model}
 \langle n_{\tilde{I}, \mathrm{far}} \rangle = \sqrt{
 \langle n_{\tilde{I}, \mathrm{elec}} \rangle^2 + p_1 I_{\mathrm{fl}} + p_2  I_{\mathrm{fl}}^2
 }.
\end{equation}
The second and third terms represent the photon shot noise $\langle n_{\tilde{I}, \mathrm{psn}} \rangle$ and fluorescence intensity noise $\langle n_{\tilde{I}, \mathrm{int}} \rangle$, respectively.
This noise model well describes the data, as shown by the black solid curve in Fig.~\ref{fig: noise}(a).
The fitted parameters were $p_1 = (5.0 \pm 0.6) \times 10^{-19}~\mathrm{A/Hz}$ and $p_2 = (5.0 \pm 0.5) \times 10^{-17}~\mathrm{/Hz}$.
The black dashed curve is the sum of $\langle n_{\tilde{I}, \mathrm{elec}} \rangle$ and the calculated shot noise given by
\begin{equation}
\label{eq: psn}
\sqrt{
\langle n_{\tilde{I}, \mathrm{elec}} \rangle^2  + 
2 \times 2 q_e I_{\mathrm{fl}}
}
,
\end{equation}
where $q_e=1.6\times 10^{-19}~\mathrm{C}$ is the elementary charge.
The factor of 2 for the shot noise term was introduced because the shot noise at the two photodiodes was assumed to be independent.
The measured shot noise coefficient $p_1 = (5.0 \pm 0.6) \times 10^{-19}~\mathrm{A/Hz}$ is close to the calculated value of $2 \times 2 q_e = 6.4 \times 10^{-19}~\mathrm{A/Hz}$.
The intensity noise $\langle n_{\tilde{I}, \mathrm{int}} \rangle = \sqrt{p_2 I_{\mathrm{fl}}^2}$ is equivalent to $\langle n_{\tilde{I}, \mathrm{psn}} \rangle$ at the fluorescence photocurrent $I_{\mathrm{fl, eqv}} = p_1 / p_2 = 10 \pm 1.6~\mathrm{mA}$.
The photon shot noise surpassed the laser intensity noise at a low fluorescence photocurrent ($< I_{\mathrm{fl, eqv}}$).

The sensor noise $n_{B}$ in the magnetic field measurement depends on demodulated photocurrent noise $n_{\tilde{I}}$ and zero-crossing slope $d\tilde{I}/d\delta$ as $n_{B} = n_{\tilde{I}} / (\gamma_e d\tilde{I}/d\delta)$.
The fluorescence dependence of the slope was measured, as shown in Fig.~\ref{fig: noise}(b).
The error bars are the estimated standard deviations of the slope; they are much smaller than the marker size.
Here, the modulation parameters and powers of the MWs and RF waves were fixed over all measurements; they were tuned at $I_{\mathrm{fl}}=7.2~\mathrm{mA}$ to maximize the slope.
Note that the optimal parameters and power depend on the incident laser power \cite{Dre11, Jen13}.
Nevertheless, we confirmed that tuning these values resulted in an improvement in the slope of about 3\% at $I_{\mathrm{fl}}=29.3~\mathrm{mA}$. We thus assumed that the relative uncertainty of the data was several percent in this measurement.

We found that the slope saturated as the incident laser power increased.
This saturation could be explained by a charge state conversion of NV centers, which led to a decrease in the contrast of a CW-ODMR peak, because [$\mathrm{N_{s}^{0}}$] was only about twice as large as [$\mathrm{NV^{-}}$] in our diamond \cite{Man05b, Col02a, Law98a}. 
A detailed investigation of this saturation is beyond the scope of this work.
The magnetic field noise density $\langle n_{B, \mathrm{far}} \rangle$ expected from the measured $ \langle n_{\tilde{I}, \mathrm{far}} \rangle$ and $d\tilde{I}/d\delta$ did not monotonically decrease as $I_{\mathrm{fl}}$ increased, as shown in Fig.~\ref{fig: noise}(c), because of the saturation of the slope.
The error bars indicate the uncertainties computed from a relative uncertainty of 5\% in the slope and the covariance matrix used in the curve fitting to $\langle n_{\tilde{I}, \mathrm{far}} \rangle$ with Eq.~(\ref{eq: noise model}).
The photocurrent dependence of $\langle n_{B, \mathrm{far}} \rangle$ suggests that good sensitivity to a magnetic field can be achieved at $I_{\mathrm{fl}}$ from 5 to 20~mA.

\subsection{%
\label{ssec: nsd}
Magnetic field noise spectral density and sensitivity
}

\begin{figure}
\includegraphics{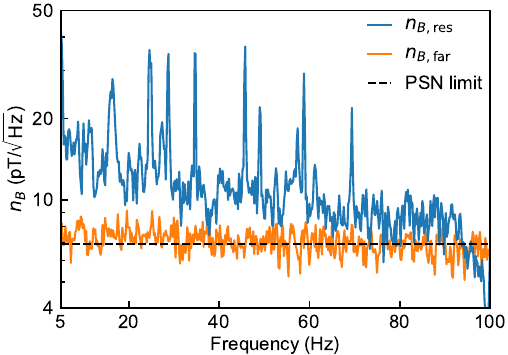}
\caption{\label{fig: nsd}
	 Single-sided noise amplitude spectral density in magnetic field measurement.
	 The blue and orange curves are the sensor noise measured with resonant and far-detuned MWs, respectively.
	 Calculated photon-shot-noise-limited sensitivity of $6.9~\mathrm{pT/\sqrt{Hz}}$ is indicated by the dashed line. PSN: photon shot noise.
}
\end{figure}

We measured single-sided noise amplitude spectral density $n_{B, \mathrm{res}}$ in a magnetic field measurement where the MWs were resonant with the central CW-ODMR peak ($\delta = 0$).
The noise spectrum $n_{B, \mathrm{res}}$ was computed using the discrete Fourier transform from a measured time trace of $\tilde{I}$ for 5~s with sampling frequency $F_{s}=400~\mathrm{Hz}$.
Figure~\ref{fig: nsd} shows the measured $n_{B, \mathrm{res}}$, which was averaged over 10 time measurements, at $I_{\mathrm{fl}}=6.4~\mathrm{mA}$ (blue solid curve).
The optimal power of the reference light was estimated from the CW-ODMR peak contrast of 3\% and a reference light power that had been optimized with the far-detuned MWs.
The zero-crossing slope $d\tilde{I}/d\delta$ was $332\pm 0.7 ~\mathrm{pA/Hz}$.
Note that the displayed $n_{B, \mathrm{res}}$ was digitally filtered by narrow-band notch filters for the harmonics of a 50-Hz power line and a band-pass filter with 3-dB cutoff frequencies of 5 and 100~Hz, which corresponds to the bandwidth of the target object (e.g., brain of a living animal).
The noise-equivalent power bandwidth $f_{\mathrm{NEP}}$ of the digital filtering was numerically calculated to be 91.9~Hz.
In this numerical calculation, white noise with a standard deviation of $\sigma$ was numerically computed with sampling frequency $F_{s}$ and digitally filtered.
The standard deviation $\sigma'$ of the filtered noise, given by $\sigma' = \sigma \sqrt{f_{\mathrm{NEP}} / (F_{s}/2)}$ \cite{Sch18}, was numerically calculated to yield $f_{\mathrm{NEP}}$.

The achieved noise density indicated a very low floor in the single-sided spectrum in the near-dc range.
The lowest noise density floor, about $9~\mathrm{pT/\sqrt{Hz}}$, was measured near 40~Hz and from 70 to 90~Hz.
The sudden drop in $n_{B, \mathrm{res}}$ at 90~Hz was due to the digital band-pass filter.
A low noise density of $15~\mathrm{pT/\sqrt{Hz}}$ was obtained even near 5~Hz, even though magnetic field noise generally deteriorates at lower frequency \cite{Zha21, Gra23, Ara22, Bar16, Bar23, Fes20}.
We attributed the noise peaks around 25~Hz to the mechanical vibration of the sensor head since these peaks shifted to lower frequencies when the sensor head was additionally supported.
The noise spectral density $n_{B, \mathrm{far}}$ measured with the far-detuned MWs at $I_{\mathrm{fl}}=6.4~\mathrm{mA}$ is shown by the orange trace in Fig.~\ref{fig: nsd}.
We found that $n_{B, \mathrm{far}}$ reached the photon-shot-noise-limited sensitivity of $6.9~\mathrm{pT/\sqrt{Hz}}$ (black dashed line).

The sensitivity of our sensor was evaluated from the average power of the measured noise, which was digitally filtered.
The sensitivity $\eta$ is defined as $\eta = \delta B \sqrt{T}$, where $\delta B$ is the minimum detectable magnetic field for measurement time $T$.
In this definition, the noise spectrum is assumed to be frequency-independent (white noise).
The standard deviation of $128\pm 2~\mathrm{pT}$ in the measured time trace is considered to represent $\delta B$ for measurement time $T=F_{s}^{-1}=2.5~\mathrm{ms}$.
Since the bandwidth of the digital band-pass filter was narrower than the measurement bandwidth $F_{s} / 2$ and the lock-in amplifier's bandwidth, $f_{\mathrm{NEP}}$ for the digital filtering was substituted for the measurement bandwidth; that is, the sensitivity was equivalent to $\eta = \delta B / \sqrt{2 f_{\mathrm{NEP}}}$ \cite{Bar16, Sch18}. 
We achieved a sensitivity of $\eta = 9.4 \pm 0.1~\mathrm{pT/\sqrt{Hz}}$.

\subsection{%
\label{ssec: adev}
Allan deviation
}

\begin{figure}
\includegraphics{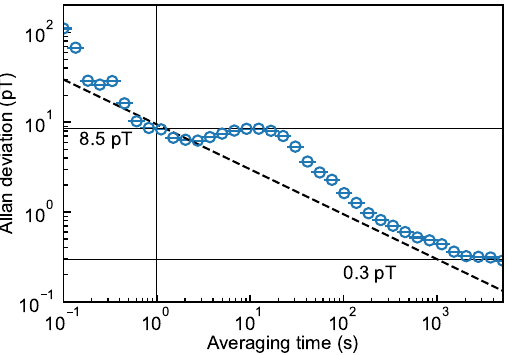}
\caption{\label{fig: adev}
	 Allan deviation as function of averaging time.
	 Open circles show calculated overlapping Allan deviation from a continuous measurement for 200 minutes.
	 Estimated uncertainties of the Allan deviations are indicated by the error bars, which are much smaller than the marker circle size.
}
\end{figure}

The Allan deviation of the noise for a measurement of about 200~minutes was computed to evaluate the stability of our sensor.
We continuously tuned the MW carrier frequency to the resonance using the demodulated photocurrent $\tilde{I}$ output, which was low-pass-filtered with a cutoff frequency of 10~Hz.
The bandwidth of the feedback response was approximately 2~Hz.
The measured noise was recorded every minute on a computer.
The notch and band-pass digital filters used in the sensitivity analysis (see Sec.~\ref{ssec: nsd}) were not used in this analysis.
We then computed the overlapping Allan deviation, as shown in Fig.~\ref{fig: adev}.
The open circles indicate Allan deviations for a given averaging time.
The error bars represent the standard deviations of the Allan deviations; they are much smaller than the marker size.

We found that the 1-second interrogation yielded an Allan deviation of 8.5~pT, which is consistent with the evaluated sensitivity ($\eta = 9.4~\mathrm{pT/\sqrt{Hz}}$).
The Allan deviation showed a bump around the 10-second averaging time.
This bump may arise from a periodic fluctuation of several tens of seconds.
We found that the bump could be suppressed to some extent by removing a 0.025-Hz component using a digital notch filter.
However, the cause of this fluctuation was not identified.
The dashed line shows the minimum detectable magnetic field $\delta B$ predicted by $\delta B = \eta / \sqrt{T}$ with the achieved sensitivity of $\eta = 9.4~\mathrm{pT/\sqrt{Hz}}$. It indicates that the Allan deviation scaled to $T^{-1/2}$ and was close to $\delta B$ at averaging times of 100 to 1000 seconds.
The Allan deviation reached 0.3~pT for an averaging time of a few thousand seconds and then seemed to saturate.
The zero-crossing slopes before and after this Allan deviation measurement were found to be almost the same.
We thus conclude that our sensor remained stable and could measure a magnetic field with a sensitivity of $\eta = 9.4 \pm 0.1~\mathrm{pT/\sqrt{Hz}}$ for at least 200 minutes.

\section{%
\label{sec: diss}
Discussion
}

The demonstrated sensitivity of $9.4 \pm 0.1~\mathrm{pT/\sqrt{Hz}}$ is the best reported value for diamond quantum sensors based on the CW-ODMR of an ensemble of NV centers \cite{Bar16, Sch18, Zha21, Gra23} in the frequency range of 5 to 100~Hz.
The previous best sensitivities were around $20$ to $30~\mathrm{pT/\sqrt{Hz}}$ in the low frequency range \cite{Zha21, Gra23} and $15~\mathrm{pT/\sqrt{Hz}}$ in the relatively high frequency range of 80~Hz--3.6~kHz \cite{Bar16}.
Moreover, in our study, the noise floor of $n_{B}$ stayed below $20~\mathrm{pT/\sqrt{Hz}}$ even at 5~Hz, as shown in Fig.~\ref{fig: nsd}.
Given that the noise environment is generally cleaner at higher frequency, the very low noise floor of about $9~\mathrm{pT/\sqrt{Hz}}$ will continue into the kilohertz range if we use a higher cutoff frequency for the lock-in amplifier's low-pass filter.
The Allan deviation analysis showed that our diamond magnetometer can interrogate a magnetic field for a long time with remarkable sensitivity.
Therefore, our sensor is capable of detecting a repetitive biomagnetic field, for example, a stimulus-evoked field, with a strength on the order of 1~pT by accumulating the signals.

CW-ODMR-based magnetometry has advantages over pulsed-MW-based magnetometry for practical applications; it has a simpler experimental setup and looser requirements for the inhomogeneities of the bias magnetic field and MWs.
Additionally, the use of a single orientation of NV center axes in our magnetometer leads to a lower requirement for the bias field alignment compared with that for multiple orientations \cite{Bar16, Gra23, Bar23}.
The sensor head design, which can approach the target object to a distance of 1~mm, relies on a simple setup and reduced requirements for the bias field.
The simplified geometry between the single orientation of NV centers and the magnetic field to be measured also facilitates various practical applications.
We note that a better sensitivity of around $2~\mathrm{pT/\sqrt{Hz}}$ in the low-frequency range (from 10~Hz), achieved using the Ramsey method, has been recently reported \cite{Bar23}; however, our sensor is more suitable for practical applications such as biomagnetic field measurement because of its simplified setup and short measurement distance.

We attributed a major part of the sensitivity improvement in this work to the long dephasing time of $T_{2}^{\ast} \sim 2~\mathrm{\mu s}$, achieved by decreasing the concentration of $^{13}$C to about 500~ppm and using a relatively low initial nitrogen concentration of 5.6~ppm.
The narrow linewidth of a CW-ODMR peak due to the long $T_{2}^{\ast}$ resulted in a high response signal to a magnetic field of $\gamma_{e}(d\tilde{I}/d\delta) = 9.3~\mathrm{A/T}$, even with the use of only a single crystallographic orientation of NV centers.
The photon-shot-noise-limited sensitivity is comparable to previously reported values \cite{Bar16, Gra23}.
In addition, the approximately five-fold improvement in the intensity noise reduction in our balanced detection over the balanced detection reported in a previous study \cite{Gra23} contributed to the good sensitivity.

A reduction in the relative intensity noise of a laser can enhance sensitivity.
The relative intensity noise of our laser (Coherent Verdi G5) was measured to be about $-130~\mathrm{dBc/Hz}$ at a modulation frequency of 6.2~kHz; the typical estimated relative intensity noise for state-of-the-art solid-state lasers at the same frequency is $-140~\mathrm{dBc/Hz}$ \cite{Ver12, Ver14}.
Therefore, a 10-dB improvement in $n_{\tilde{I}, \mathrm{int}}^{2}$ is feasible. This would result in a photon-shot-noise-limited sensitivity at up to $I_{\mathrm{fl}}\sim 100~\mathrm{mA}$.

It is expected that the zero-crossing slope can be increased by extending the dephasing time $T_{2}^{\ast}$ for the diamond.
For example, a very long dephasing time of $8.5~\mathrm{\mu s}$ with $\mathrm{[NV^{-}]} = 0.4~\mathrm{ppm}$ has been reported \cite{Zha21}.
This long dephasing time will offer a four-fold improvement if the same fluorescence intensity is available since the shot-noise-limited sensitivity is proportional to the linewidth of a CW-ODMR peak \cite{Dre11, Bar16}.
Although the lower $[\mathrm{NV^{-}}]$ emits weaker fluorescence, a photocurrent of up to 10~mA can be obtained by increasing the incident laser power.
In addition, the fluorescence collection efficiency can be boosted to approximately unity by using a total internal reflection lens and a light pipe \cite{Als23, Bar23}.

\section{%
\label{sec: concl}
Conclusions
}

We demonstrated a sensitive diamond magnetometer with a magnetic field sensitivity of $9.4 \pm 0.1~\mathrm{pT/\sqrt{Hz}}$ in a near-dc frequency range of 5 to 100~Hz.
The magnetometer can closely approach the target object and the measurement distance from the sensing volume was about 1~mm.
The Allan deviation indicated that our magnetometer can measure magnetic fields of 8.5 and 0.3~pT with a unity signal-to-noise ratio by interrogating for 1 second and several thousands of seconds, respectively.
Our high-sensitivity diamond magnetometer was designed to be compatible with practical applications, including the measurement of the encephalomagnetic field of a living animal.
The sensitivity improvement achieved in this work is an important step toward realizing magnetoencephalography under ambient conditions with millimeter-scale spatial resolution.

\begin{acknowledgments}
This work was supported by the MEXT Quantum Leap Flagship Program (MEXT Q-LEAP) Grant No. JPMXS0118067395 and JPMXS0118068379.
\end{acknowledgments}

\bibliography{2023PRAppl}

\end{document}